\documentclass[lettersize,journal]{IEEEtran}
\usepackage{amsmath,amsfonts}
\usepackage{algorithmic}
\usepackage{algorithm}
\usepackage{array}
\usepackage[caption=false,font=normalsize,labelfont=sf,textfont=sf]{subfig}
\usepackage{textcomp}
\usepackage{stfloats}
\usepackage{url}
\usepackage{verbatim}
\usepackage{graphicx}
\usepackage{cite}
\begin{document}
\title{Design of Ultrafast All-Optical Pseudo Binary Random Sequence Generator, $4$-bit Multiplier and Divider using $2\times2$ Silicon Micro-ring Resonators}

\author{Aayushman Ghosh,~%~\IEEEmembership{Student Member,~IEEE,}
        Sayan~Sarkar,~%~\IEEEmembership{Student Member,~IEEE,}
        and Sukhdev Roy%~\IEEEmembership{Senior Member,~IEEE}% <-this % stops a space
\thanks{The work of Aayushman Ghosh was carried out at the Department of Physics and Computer Science, Dayalbagh Educational University under the Summer Research Fellowship programme sponsored by the Indian Academy of Sciences. Sukhdev Roy acknowledges the University Grants Commission, India [F.530/14/DRS-III/2015(SAP-I)]; and Department of Science and Technology, India [CRG/2021/005139 and MTR/2021/000742], for partial finding of this work.}
\thanks{Aayushman Ghosh is with the Department of Electronics and Telecommunication Engineering, Indian Institute of Engineering Science and Technology, Shibpur 711103, India, e-mail: (510719076.aayushman@students.iiests.ac.in).}% <-this % stops a space
\thanks{Sayan Sarkar is with Department of Electronics and Computer Engineering, The Hong Kong University of Science and Technology, Hong Kong, e-mail: (ssarkar@connect.ust.hk).}
\thanks{Sukhdev Roy is with the the Department of Physics and Computer Science, Dayalbagh Educational Institute, Agra 282005, India, e-mail: (sukhdevroy@dei.ac.in).}}% <-this % stops a space


\markboth{Manuscript Submitted at Optik (2022)}
{Ghosh\MakeLowercase{\textit{et al.}}: Bare Demo of IEEEtran.cls for IEEE Journals}



\maketitle
\vspace{-10 cm}
\begin{abstract}
All-optical devices are essential for next generation ultrafast, ultralow-power and ultrahigh bandwidth information processing systems. Silicon microring resonators (SiMRR) provide a versatile platform for all-optical switching and CMOS– compatible computing, with added advantages of high Q-factor, tunability, compactness, cascadability and scalability. A detailed theoretical analysis of ultrafast all-optical switching in $2\times2$ SiMRRs has been carried out incorporating the effects of two photon absorption induced free-carrier injection and thermo optic effect. The results have been used to design simple and compact all-optical $3$-bit and $4$-bit pseudo-random binary sequence generators and the first reported designs of all-optical $4\times4$-bit multiplier and divider. The designs have been optimized for low-power ($\sim 28$ mW), ultrafast ($\sim 22$ ps) operation with high modulation depth ($80\%$), enabling logic operations at $45$ Gb/s.
\end{abstract}


% \fi
\IEEEpeerreviewmaketitle

\section{INTRODUCTION}
\IEEEPARstart{T}{he} tremendous growth in internet data traffic has led to an increasing demand for ultrafast, ultrahigh bandwidth and more energy-efficient information processing technology. This has prompted intense research on ultrasensitive modulators [1–4]. Traditional state-of-the art electro-optic modulators (EOMs) [5] based on carrier-injecting effect [6], thermo-optic effect [7], and Pockels effect [8] have recorded significant progress recently. EOMs designed on silicon [9] and thin-film $LiNbO_3$ platforms [10] have demonstrated modulation bandwidth close to $100$ $GHz$ with modulation speed over $50$ $Gb/s$. Nevertheless, the intrinsic limitation from parasitic capacitance and resistance in the electric circuit, along with the energy consumption from electro-optic conversion still limit their performance in on-chip applications [5-11]. Although all-optical modulators (AOMs) that use a pump light to control a probe light can overcome the EOMs bottleneck [11-27], AOMs could eliminate the electro-optic conversion, enable parallel processing, achieve lower power consumption with a larger bandwidth [14], and open the path to ultrafast on-chip optical connection networks as well as integrated photonic devices [1–4, 11–27]. 

Optical switches (OS) are essential building components of AOMs that can realize various optical logic operations [15–16], wavelength conversion [17], and quantum photonic circuit applications [18]. Recently, all-optical logic devices have been designed for programmable logic [19], nanowire networks [20] and optical computing [21]. Hence, studying complex logic functions in all-optical regime is important for next-generation computing. Traditional schemes of OS varied over the years, starting from inducing resonance control in the cavity via a pump pulse [22] in nonlinear optical fibers [23]. More recent techniques include semiconductor optical amplifiers (SOA) [24], Mach-Zehnder Interferometer (MZI) [25], silicon microring resonators (SiMRR) [15–17], photonic crystals [26], surface plasmon polaritons [4], and metamaterials [27]. SOA and MZI based conventional logic designs lack on-chip integration capability due to bulky dimension and flexibility [28]. Although the footprint of plasmonic logic gates is smaller than the wavelength scale, the large optical loss makes them difficult to use in integrated chip [29]. In contrast, MRRs fabricated with silicon wire waveguide have attracted considerable attention due to its CMOS compatible fabrication with advantages of low-loss, tight confinement of guided modes, compact footprint and optical field enhancement. This enables SiMRRs as an optimal choice for integrated chips [30]. SiMRRs have been used to realize all-optical logic operations that include Boolean AND, OR, NOT, XOR logic gates, re-configurable logic circuits, flip-flops and arithmetic logic units [15, 31–33]. To the best of our knowledge, complex logic operations such as, multipliers and dividers using SiMRRs have not been demonstrated as yet. 

Multipliers are basic building blocks of any digital signal processing system for frequency multiplication, multiply-accumulate unit and fast-fourier transform [34]. Hence, a cost-effective implementation of all-optical multiplier is important for integration into Photonic Integrated platforms (PICs). Designs of $2\times2$–bit optical multipliers have been reported using EOMs [35], Programmable Logic Devices [36], SOA based MZIs [34] and canonical logic units-based programmable logic arrays leveraging the effect of four-wave mixing [37]. These designs cannot be practically realized in PICs due to their large footprint, and cannot be generalized for higher number of bits. Similarly, dividers find application in forward error detection and correction (FEC) along with DSP systems [38]. Aikawa \textit{et al.,} demonstrated an all-optical divider based on a single SOA-MZI configuration for all-optical FEC that enables operation $\sim 40$ $Gb/s$ [38]. Another important complex logic operation is the generation of Pseudo Random Binary Sequences (PRBS), which is useful in multiple domains ranging from pattern recognition, cryptography to code division multiple access communication systems [39]. Recently, Rakshit \textit{et al.,} suggested an all-optical ultrafast PRBS system using MRRs [39], which can further be optimized to use minimal rings to reduce the footprint. 

\begin{figure*}
    \centering
    \includegraphics[width = 15.8cm, height = 5.5cm]{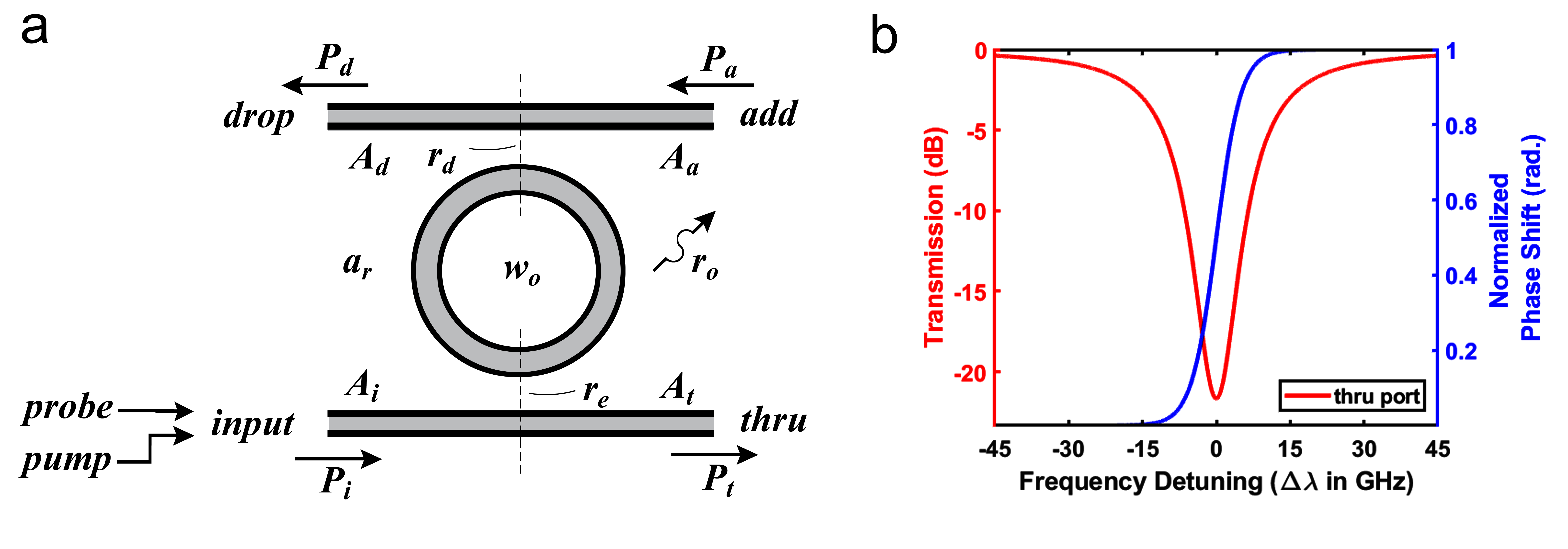}
    \caption{(a) Schematic representation of a ($2\times2$) add-drop Silicon Microring Resonator. (b) Low power Transmission spectra (red) of the $Si$MRR, collected at the thru port, simulated using experimental conditions [43]. The normalized phase shift of the MRR is shown in blue.}
    \label{fig1}
\end{figure*}

In this paper, we propose novel designs of $4\times4$-bit All-Optical Multiplier and Divider circuit, along with $3-$bit and $4-$bit PRBS Generator using $2\times2$ add-drop MRRs. The designs were numerically simulated using a variable order Adams-Bashforth predictor corrector method that is based on nonlinear coupled-mode theory. All-optical switching has been further optimized to realize low-power ultrafast logic operations near telecommunication wavelengths, making them suitable for all-optical signal processing.

\section{THEORETICAL MODEL}

We theoretically analyze all-optical switching in silicon ring resonators considering TPA, FCA-FCD and the optical Kerr effect [31-33, 40]. TPA ensures carrier generation, which induces more optical loss through FCA, and the free carrier dispersion (FCD) effect shifts the resonance wavelength to shorter peaks [40–41]. Logic operations using ring resonators depend on rapidly varying the effective index for all-optical intensity modulation. Due to weak Kerr effect in silicon at telecommunication wavelengths, the free-carrier plasma dispersion effect is the preferred mechanism for the effective-index change [40]. In this paper, all-optical switching is studied using a $(2\times2)$ add-drop silicon MRR as shown in Fig. 1. $P_i$ and $P_a$ are the input and add ports, whereas $P_t$ and $P_d$ are through and drop ports. A low-powered probe pulse at a non-resonant wavelength close to the edge of the resonance spectrum is applied, which is transmitted to $P_d$ in the absence of the pump pulse. The arrival of the high-powered control light (pump) induces free carriers in the MRR due to TPA, which reduces the refractive index of silicon through FCD and involves a blue shift in the resonance spectrum [41]. The resulting change in the resonance couples with the probe and yields a high transmission at $P_t$ and a relatively low transmission at $P_d$. Once the control light passes, the resonant wavelength and probe transmission relax back due to fast recombination of the injected free carriers. The time interval depends on the free-carrier recombination lifetime on the SOI interface [31–33].
%\vspace{-1mm}
%\subsection{$\chi^{(3)}$ Dynamics of $Si$ MRR Cavity} 
The time-domain nonlinear coupled-mode theory can be used to accurately describe the dynamic behavior inside the cavity (SiMRR) [40–44], [31–33]. A pair of waveguides are coupled to the ring resonator for the propagation of probe and pump pulses. As is evident from Fig. 1, $A_i$, $A_a$, $A_t$, and $A_d$ denote the field amplitudes at ports $P_i$, $P_a$, $P_t$, and $P_d$ respectively. For mathematical simplification, the cavity is represented as a lumped oscillator where, $|A_x|^2 (x = i, a, t, d)$ denotes the power of the waveguide mode that supports a travelling wave $A_r(t)$ $(A_i \neq0, A_a=0)$. Hence, $|A_r(t)|^2$ represents the total power flowing through the cross-section $A$ of the ring waveguide at any time $t$. Hence, the total energy stored in the ring is $|a_r(t)|^2$ having an amplitude $a_r(t)$. The stored energy and power flow in the ring are related through $|a_r(t)|^2 = |A_r(t)|^2 2\pi R/v_g$ where, $2\pi R = L$ is the round-trip length of the ring (cavity). Here, $R$ is the ring radius, and $v_g$ is the group velocity. The ring resonator in Fig. 1 is analyzed further by considering the resonant frequency to be $\omega_o$ and an amplitude decay-constant that follows $r = r_o + r_e + r_d$. Here, $r_o = \alpha_ov_g/2$ denotes the power lost due to intrinsic effects such as linear absorption and rough sidewall scattering, $r_e = \mu^2/2 = \kappa_e^2v_g/2$ is the power lost due to external coupling with the input waveguide and $r_d = \mu_d^2/2 = \kappa_d^2v_g/2$ is the decay rate due to external coupling between the ring and drop waveguide. $\tau$ is the total decay constant, whereas $\mu_e$ and $\mu_d$ are defined as the mutual coupling coefficients and $\kappa_e$, $\kappa_d$ as the corresponding power coupling coefficients. From the energy flow considerations, the rate of change of ring energy is expressed as,
\begin{equation}
    \frac{da_r}{dt} = \left(j\omega_o - \frac{1}{\tau}\right)a_r - j\mu_eA_i 
\end{equation}

Considering a small round-trip loss ($\alpha_oL<< 1$), weak coupling condition ($\kappa^2<<1$) and small frequency variation $\omega - \omega_o << \omega_o$, the amplitude of field coupled into the MRR waveguide can be written as [31-33],
\begin{equation}
    a_r = \frac{-j\sqrt{2r_e}}{j(\omega - \widetilde{\omega_o}) + r}A_i
\end{equation}
The output field amplitude in the waveguide is given by
\begin{subequations}
\begin{gather}
A_t(t)=A_i(t)-j \sqrt{2 r_e} a_r(t) \\
A_d(t)=-j \sqrt{2 r_d} a_r(t)
\end{gather}
\end{subequations} 
By substituting (2) in (3a) and (3b), the transmission response at ports $P_t$ and $P_d$ are written as

\begin{subequations}
\begin{gather}
    \mathcal{T}(\omega) = \frac{\mathcal{A}_t}{\mathcal{A}_i} = \frac{j(\omega - \widetilde{\omega_o}) + r - 2r_e}{j(\omega - \widetilde{\omega_o}) + r} \\
    \mathcal{D}(\omega) = \frac{\mathcal{A}_d}{\mathcal{A}_i} = \frac{-2\sqrt{r_er_d}}{j(\omega - \widetilde{\omega_o)} + r}
\end{gather}
\end{subequations}

where, the complex resonance frequency can be written as $\widetilde{\omega_o} = \omega_o + jr_o$. The decay constant $r$ is defined as $r = \omega_o/2Q_o + \omega_o/2Q_e + \omega_o/2Q_d$, where $Q_o$, $Q_e$ and $Q_d$ ($Q^{-1} = Q_o^{-1} + Q_e^{-1} + Q_d^{-1}$) are the corresponding quality factors. Under critical coupling conditions, the total signal power is extracted by the resonator and the total microring loss is equal to the coupling loss ($r_e = r_d + r_o$) [40]. The full-width at half maximum (FWHM) bandwidth of the Gaussian responses in (2)-(4) is $\Delta\omega_{FWHM} = 1/\tau_p$. The transmission spectrum of the resonant cavity is extremely sensitive to the refractive index of the ring waveguide. Therefore, efficient amplitude modulation is possible by changing the refractive index of the ring resonator $(\Delta n)$. A small change in the refractive index $\Delta\Tilde{n} = \Delta n - jc\Delta\alpha/\omega_o$, where $\alpha$ is the material absorption, results in a small
shift of the complex resonance frequency $\widetilde{\Delta\omega_o} = \Delta\omega_o + j\Delta r$. To achieve ultra-fast switching, TPA-induced FCD is used to modulate the refractive index of silicon. The electric field in the ring waveguide is expressed as $\Vec{E} = a_r\Vec{e}(x,y)e^{-j\Tilde{\beta}z}$, with energy normalization defined as $1/2\varepsilon_o\iiint n^2|\Vec{E}|^2dxdydz = |a_r|^2$. Here, $\Tilde{\beta}$ denotes the complex propagation constant, and $z$ is defined along the ring circumference whereas $x$ and $y$ in the cross-section of the ring waveguide. Since, $(\alpha_oL << 1)$, we can make approximations $\iiint dxdydz \approx L\iint dxdy$. Considering $n$ and $\Delta\Tilde{n}$ to be uniform in the waveguide core, the real and imaginary part responsible for the resonance frequency shift becomes,
\begin{equation}
    \Delta\omega_o^L = -\omega_o\Gamma\frac{\Delta n}{n_r} \hspace{1.5 mm}\text{and}\hspace{1.5mm} \Delta r_o^L = \Gamma\frac{c\Delta\alpha}{n_r}  
\end{equation}
Here, $n_r$ is the value of $n$ in the core of the ring waveguide and $\Gamma$ is the confinement factor. The induced a (linear) change in the refractive index and in the absorption coefficient ($\Delta n$ and $\Delta\alpha$) at $1.55$ $\mu m$ is expressed as [40–44],
\begin{equation}
\begin{aligned}
\Delta n=\Delta n_e+\Delta n_h &=-\left[8.8 \times 10^{-22} \Delta N_e\right.\\
&\left.+8.5 \times 10^{-18}\left(\Delta N_h\right)^{0.8}\right] \\
\Delta \alpha=\Delta \alpha_e+\Delta \alpha_h &=\left[8.5 \times 10^{-18} \Delta N_e\right.\\
&\left.+6 \times 10^{-18} \Delta N_h\right]
\end{aligned}
\end{equation}
where $\Delta n_e$ and $\Delta n_h$ are the corresponding changes in refractive index due to change in electron concentration $(cm^{-3})$ $\Delta N_e$ and hole concentration $\Delta N_h$ respectively, whereas $\Delta\alpha_e$ $(cm^{-1})$ and $\Delta\alpha_h(cm^{-1})$ are absorption coefficient variations due to $\Delta N_e$ and $\Delta N_h$. Operation around the telecommunication $C$-band $(1550$ $nm)$ in silicon allows excitation of free carriers with photon energies well below the band-gap. This further induces small nonlinear index and absorption changes $\Delta N^{NL}$ and $\Delta\alpha^{NL}$ due to the Kerr effect, which are given by, $\Delta n^{NL} = n_{2I}I_{pump}$ and $\Delta\alpha^{NL} = \beta_{TPA}I_{pump}$, where $n_{2I} = 0.45\times 10^{-13}$ $cm^2/W$ and $\beta_{TPA} = 0.81$ $cm/GW$ are the respective Kerr and TPA coefficients at $1.55$ $\mu m$ for Silicon $Si\langle110\rangle$ orientation. The value of $(n_{2I})$ nearly remains the same for $Si\langle111\rangle$ orientation, but $(\beta_{TPA})$ changes due to their fundamental change in crystallographic direction [40]. Here, the pump intensity coupled into the ring $I_{pump} = c\varepsilon_o n|\overrightarrow{E_{pump}}|^2/2$, where $|\overrightarrow{E_{pump}}|$ has the same form of electric field in the ring waveguide core with amplitude $a_r^{pump}$. Therefore, the effect of nonlinear changes to the complex resonance wavelength is expressed as,
\begin{subequations}
    \begin{gather}
        \Delta\omega_o^{NL} = -\omega_o\frac{cn_{2I}}{n_r^2}\frac{|a_r^{pump}|^2}{V_{eff}} \\
        \Delta r_o^{NL} = \beta_{TPA}\frac{c^2}{2n_r^2}\frac{|a_r^{pump}|^2}{V_{eff}}
    \end{gather}
\end{subequations}
where. $V_{eff} = LA_{eff}$ is the effective volume of the cavity. $A_{eff}$ is defined as the effective area. A small red shift in resonance is caused due to weak Kerr coefficient and $\Delta n^{NL} > 0$. However, the blue shift caused by TPA dominates as $|\Delta\omega_o^{NL}|$ $<< |\Delta\omega_o^L|$. The change in free carrier concentration due to TPA-induced FCD ($\Delta N_e = \Delta N_h = N_{fc}$) is given by,
\begin{equation}
    \frac{dN_{fc}(t)}{dt} = \frac{1}{\hbar\omega_{pump}}\Delta r_o^{NL}\frac{|a_r^{pump}|^2}{V} - \frac{N_{fc}(t)}{\tau_{fc}}
\end{equation}
where $\hbar\omega_{pump}$ is the energy due to high-powered pump. $\tau_{fc}$ is defined as the free carrier recombination lifetime, which was assumed to be a fixed value that does not change with carrier density. The temperature evolution inside the cavity termed as the Thermo-Optic Effect (TOE) can similarly be modeled [44]. TPA, FCA and linear absorption are generally the cause for the heat generated inside the ring. This shifts the resonance wavelength to higher peaks and induces a red shift. The generated heat inside the waveguide generally diffuses through conduction and hence a thermal decay time $\tau_{th}$ is used to describe the temperature evolution as,
\begin{equation}
\begin{aligned}
\frac{d\Delta T}{dt} = \frac{a_r^{pump}}{\rho_{Si}c_{Si}V_{eff}}&\left(\frac{r_oc}{n_r} + \frac{a_r^{pump}\beta_{TPA}c^2}{n_r^2V_{TPA}} \right.\\
&\left. + \frac{\sigma_{FCA}N_{fc}c}{n_r}\right) - \frac{\Delta T}{\tau_{th}}
\end{aligned}
\end{equation}
where, $\rho_{Si}$ and $c_{Si}$ are the density and the constant-pressure specific heat capacity of silicon, and $\sigma_{FCA}$ is the corresponding FCA coefficient. The proposed nonlinear system has three specific time constants $\tau_p$, $\tau_{fc}$, and $\tau_{th}$. The switching speed is determined by the magnitude of the decay constants and is affected by the charging time and the carrier recombination time. Generally, $\tau_{fc}$ drives the switching time as it is longer than the cavity photon lifetime $\tau_p$. The cross-sectional geometry of the ring can be modified in such a way that the FCD blueshift compensates the dominant TOE redshift. This implies fabricating a resonant photonic structure with a long free-carrier lifetime $\tau_{fc}$ [45]. 

It is necessary to analyze this nonlinear dynamic process to understand the transient response of the cavity. Both the temporal and spatial variation of all the quantities along the ring cavity must be considered to perform such an analysis. This would complicate the analysis. To simplify it, we assume a low loss ring with round-trip time much lesser than the duration of optical pulses and express only in terms of the temporal variation of optical and material properties of the ring. The temporal evolution of the energy-normalized intracavity pump and probe fields $a_r^{pump}$ and $a_r^{probe}$ can be derived by substituting $j(\omega - \omega_{pump})\xrightarrow{}d/dt$, $j(\omega - \omega_{probe})\xrightarrow{}d/dt$ and $\Tilde{\omega}\xrightarrow{}(\omega_o + \Delta\omega_o^{L} + \Delta\omega_o^{NL} + \Delta\omega_o^{th}) + j(r_o + \Delta r_o^{L} + \Delta r_o^{NL} + \Delta r_o^{th})$ in (2). Here, $\Delta\omega_o^{th} = \omega_o\kappa_{\theta}\Delta T/n_r$ and $\Delta r_o^{th}$ $=\sigma_{FCA}N_{fc}c/2n_r$ are the real and imaginary part of the complex wavelength shift incurred due to heat generation in the cavity. To incorporate add-drop geometry, the equations are modified for $(1\times1)$ ring configuration as,
\begin{equation}
        \begin{aligned}
            &\frac{d}{dt}a_r^{pump}(t) = -j\left[\omega_{pump} - \omega_o - \Delta\omega_o^L(t) - \right.\\
            &\left. \Delta\omega_o^{NL}(t) - \Delta\omega_o^{th}(t)\right]a_r^{pump}(t) - \left[r_o + \Delta r_o^L(t) \right. \\
            &\left. +\Delta r_o^{NL}(t) + \Delta r_o^{th}(t)\right]a_r^{pump}(t) - \\ 
            & a_r^{pump}(t)(r_e + r_d) - j\sqrt{2r_e}A_i^{pump}(t) \\
        \end{aligned}
\end{equation}
\begin{equation}
        \begin{aligned}
            &\frac{d}{dt}a_r^{probe}(t) = -j\left[\omega_{probe} - \omega_o - \Delta\omega_o^L(t) - \right.\\
            &\left. \Delta\omega_o^{NL}(t) - \Delta\omega_o^{th}(t)\right]a_r^{probe}(t) - \left[r_o + \Delta r_o^L(t) \right. \\
            &\left. +\Delta r_o^{NL}(t) + \Delta r_o^{th}(t)\right]a_r^{probe}(t) - \\ 
            & a_r^{probe}(t)(r_e + r_d) - j\sqrt{2r_e}A_i^{probe}(t) \\
        \end{aligned}
\end{equation}
The same equations can be considered when $A_a \neq 0$ at $P_a$. Symmetry enables us to demonstrate a $(2\times2)$ switch using a modified version of (1). 
\begin{equation}
    \frac{da_r}{dt} = \left(j\omega_o - \frac{1}{\tau}\right)a_r - j\mu_eA_i - j\mu_dA_a
\end{equation}
where the output field amplitudes in the waveguide are expressed as,
\begin{subequations}
    \begin{gather}
        A_t(t) = A_i(t) - 2j\sqrt{2r_e}a_r(t) \\
        A_d(t) = A_a(t) - 2j\sqrt{2r_d}a_r(t)
    \end{gather}
\end{subequations}
%\vspace{2cm}
\section{RESULTS AND DISCUSSION}
Computing the complex resonance frequency and the transient response by numerically analyzing the coupled ODEs $(1)-(11)$ is a seemingly difficult task. The large order of differences $(\sim10^{34})$ in magnitude $(a_r, N_{fc}, \Delta T)$ creates additional hurdles. Hence, achieving correct dynamic results using the Runge-Kutta method seems challenging [42]. In light of these observations, there exist some strategies to solve the coupled ODEs [40–45]. Zhang \textit{et al.} have normalized the parameters $(a_r, N_{fc}, \Delta T, t)$ and used a linear stability eigenmatrix to solve ODEs for the boundaries of self-pulsation and bistability [44]. Cea \textit{et al.} have used a variable order Adams-Bashforth predictor-corrector based method to solve the nonlinear system [42]. Johnson \textit{et al.,} have adapted the same methodology [41]. Here, we also follow Cea \textit{et al.,} and use a variable order Adams-Bashforth-Moulton predictor-corrector method to analyze the coupled nonlinear system of equations. We use experimental data from [43] to validate our theoretical model. A microring resonator of $7$ $\mu m$ radius is considered, with $A = 450$ $nm\times220$ $nm$ and a coupling coefficient of $\kappa^2 = 0.063$. In the experiment, Borghi \textit{et al.} used a $cw$ laser source with $\lambda_{probe} = 1555$ $nm$, and a picosecond (pulse width$-40$ $ps$, peak power $\sim1.5$ $W$) laser pump ($\lambda_{pump} = 1550$ $nm$) with 1 MHz of repetition rate to measure the Self-Pulsation in an MRR. The low noise transmission spectra was observed, and the loaded Q-factor ($Q_{L}$) of $6.5\times10^3$ was extracted. We numerically simulated the same and presented a zoomed-in version of this transmission spectra around $1555$ $nm$ (see Fig. 1(b)). We numerically calculated an $\Delta\lambda_{FWHM}$ of $0.223$ $nm$ that results in a $Q_L$ of $6.67\times10^3$, which is farely accurate and close to what is observed experimentally. This proves that the theoretical simulations are in good agreement with the reported experimental results.     

\begin{table}[ht]
\caption{PARAMETERS USED IN SIMULATION [6, 30, 42, 43, 46]}
\centering
\begin{tabular}{l|c}
    \hline\hline 
       Microresonator Parameters & Value\\
       \hline
        Radius ($R$) & $7$ $\mu m$\\
        Rectangular Cross-section Area ($A$) & $450\times250$ $nm^2$\\

        $\lambda_{pump}(\lambda_{res1})$ & $1550$ $nm$\\
        $Q_{res1}(\lambda_{res1}/\Delta\lambda_{FWHM1})$ & $11076$\\

        $\Delta\lambda_{FWHM1}$ & $0.14$ $nm$\\
        $\tau_{cav1}(\lambda_{res1}^2/2\pi c\Delta\lambda_{FWHM1})$ & $9.1$ $ps$\\

        $\lambda_{probe}(\lambda_{res2})$ & $1555$ $nm$\\
        $Q_{res2}(\lambda_{res2}/\Delta\lambda_{FWHM2})$ & $9804$\\

        $\Delta\lambda_{FWHM2}$ & $0.16$ $nm$\\
        $\tau_{cav2}(\lambda_{res2}^2/2\pi c\Delta\lambda_{FWHM2})$ & $8.1$ $ps$\\

        Free Carrier Lifetime ($\tau_{fc}$) & $12.2$ $ps$\\
        Thermal decay time ($\tau_{th}$) & $84$ $ns$\\
        Pump pulse width & $0.1$ $ps$\\
        Probe pulse width & $18$ $ps$\\
        Pump-probe delay & $6$ $ps$\\
        $P_c(E_{pump})$ & $\sim28$ $mW$($\sim2.9$ $pJ$) \\
        $FSR = \lambda_o^2/Ln_g$ & $18$ $nm$\\
        Group index $(n_g)$ & $4.44$\\
        $\Gamma$ & $0.92$\\
        $n_r$ & $3.485$\\
        $A/A_{eff}$ & $0.82$\\
        $\beta_{TPA}$ & 0.81 $cm/GW$\\
        $n_{2I}$ & $0.45\times10^{-23}$ $cm^2/W$\\
        $\rho_{Si}$ & $2.3290$ $gcm^{-3}$\\
        $c_{Si}$ & $700$ $J/KgK$\\
        TOE Coefficient of Silicon ($dn/d\Delta T$) & $1.86\times10^{-4}$ $K^{-1}$\\
        FCA coefficient ($\sigma_{FCA}$) & $1.5\times10^{-21}$ $m^2$\\
        TPA effective volume ($V_{TPA}$) & $5.35\times10^{-18}$ $m^3$\\
        \hline\hline
    \end{tabular}
    \label{table1}
\end{table}

Moving forward, the time evolution of the Normalized (NPT) Power for Probe at the output of the silicon $(2\times2)$ cavity is studied to gain insight into the nonlinear mechanism and to realize high bit-rate logic operation. We consider the values for SiMRR parameters from reported experiments [6, 30, 42, 43, 46], listed in Table I, and optimize them to realize ultrafast all-optical switching. We still consider the cavity with $7$ $\mu m$ radius and a core dimension of $450\times250$ $nm$. Considering a $3$ $\mu m$ coupling length, the coupling coefficient ($\kappa^2$) is $\cong 0.063$ [43]. The absence of the pump yields a high transmission of $\lambda_{probe}$ at $P_d$, and a low transmission at $P_t$, whereas, in the presence of the pump, $\lambda_{probe}$ is switched to $P_t$ resulting in a low transmission at $P_d$. To ensure modulation in the ring, the applied control power ($P_c$) has to be greater than a threshold, given by $P_c^2 \geq \pi^2n_g^2n_{eff}h\nu_cV_{eff}^2/2\Gamma n_f\beta\lambda_2^2Q_2^2Q_1\tau_{fc}$ where $n_f$ is the ratio between refractive index change and carrier density, $n_{eff}$ is effective index of the ring waveguide, and $h\nu_c$ is control photon energy [6]. From the given relation it can be easily derived that $P_c$ is inversely related to both $\tau_{fc}$ and $Q$. Hence, to realize high bit-rate, ultrafast logic operation, we consider the smallest reported value of $\tau_{fc} = 12$ $ps$ [46], and Q-factor $= 11,076$ [31]. This results in $P_c = 28$ $mW$, a switch-on/off time of $9$ $ps/22$ $ps$ and a modulation of $>80\%$. An additional pump-probe delay of $6$ $ps$ is also taken into consideration to achieve a high switching contrast of $>80\%$. We consider the experimental conditions of Xu $\textit{et al.}$ [6] to confirm the performance of the proposed circuits with existing studies [31-33], with a $5$ $mW$ probe laser, out of which $\sim 30\% (1.5$ $mW)$ gets coupled through a nano-tapered waveguide [6]. During off-resonance condition $\sim90\%$ $(1.35$ $mW)$ and at resonance only $\sim20\%$ $(0.3$ $mW)$ gets coupled either from $P_i$ to $P_t$ or from $P_a$ to $P_d$. We consider a low-loss ring with an insertion loss of $\sim 3.5$ $dB$ when the optical signal is coupled to the input waveguide of the SiMRR and $\sim 6$ $dB$ loss due to scattering and absorption. To realize logic operations, the transmission spectra of the output probe power was divided into two thresholds. A threshold of $P_{OL} = 0.4$ $mW$ was considered the maximum limit of logic 0. Similarly, to realize logic 1, $P_{OH} = 0.8$ $mW$ was considered as the minimum threshold.

\section{DESIGN OF ALL-OPTICAL CIRCUITS}

The optical switching characteristics of a SiMRR, namely the variation in NPT with time, have been analyzed by solving equations (1)–(11) through computer simulations, considering reported experimental parameters in Table I [6, 30, 42, 43, 46] and the results used for designing all-optical 3-bit and 4-bit pseudo-random binary sequence generators and $4\times4$-bit multiplier and divider circuits.

\vspace{-1.5mm}
\subsection{3-bit and 4-bit Pseudo Random Binary Sequence Generator}
Fig. 2(a) and 2(f) presents novel designs of $3-$bit and $4-$bit Pseudo Random Binary Sequence (PRBS) Generator. PRBS is one of the essential digital sequential logic circuit, consisting of several clocked $D$ flip-flops in series with an XOR gate in feedback. The total number of bits generated by the PRBS influences the tap-position and number of $D$ flip-flops. For a $n-$bit PRBS the output sequence will be in a periodic manner with the length of output sequence equal to $2^n - 1$, where `$n$' is the number of $D$ flip-flops. All the SiMRRs are triggered by the same optically clocked pump signal ‘Clk’. The output of each MRR after each Clk pulse is depicted in Truth Tables II (3-bit) and III (4-bit). Fig. 2(c), (d), (e) show the output of the 3-bit implementation whereas Fig. 2(h), (i), (j) and (k) depict the output of the 4-bit PRBS.
\begin{table}[ht]
    \caption{TRUTH TABLE OF 3-BIT DEGREE PRBS GENERATOR WITH INITIAL SEQUENCE, $Q_{C}Q_{B}Q_{A}$ $\xrightarrow{}$ 001}
    \centering
    \begin{tabular}{c|c|c|c|c}
    \hline\hline
    Clock & $Q_C$ & $Q_B$ & $Q_A$ & $XOR$ \\ 
    Cycle (Clk) & (mW) & (mW) & (mW) & Output (mW)\\
    \hline
    0 (initial) & 0 & 0 & 1(1.1) & 1(1.3)\\
    $1^{st}$ & 0 & 1(1.1) & 1(1.3) & 1(1.2)\\
    $2^{nd}$ & 1(1.2) & 1(1.3) & 1(1.3) & 0\\
    $3^{rd}$ & 1(1.3) & 1(1.0) & 0 & 1(1.1)\\
    $4^{th}$ & 1(0.9) & 0 & 1(0.9) & 0\\
    $5^{th}$ & 0 & 1(0.9) & 0 & 0\\
    $6^{th}$ & 1(1.0) & 0 & 0 & 1(0.9)\\
    $7^{th}$ (repeat) & 0 & 0 & 1(1.0) & 1(0.9)\\
    \hline
    \hline
    \end{tabular}
    \label{table2}
\end{table}

Considering the implementation of a $3-$bit PRBS Generator, $M_1 - M_3$ denote the $D$ flip flops, whereas $M_4$ signifies the XOR gate. A combination of wavelength converter (WC) and waveguide amplifier (WA) is placed at the through output port of $M_3$ before connecting it to $M_4$, such that $\lambda_{probe}$ can be converted to $\lambda_{pump}$, to amplify the signal to act as a pump signal. The initial bit pattern of $Q_CQ_BQ_A\xrightarrow{}001$ is assumed for computation of the subsequent Clk cycles. Case (i): $Q_CQ_BQ_A\xrightarrow{}011$, $XOR\xrightarrow{}1$. The output of XOR is fed to the input of $M_1$ resulting in $Q_A = 1$. Due to the previous state of $Q_A = 1$, thru port of $M_2$, i.e., $Q_B$ becomes $1$. Similarly, $Q_C = 1$, due to the previous state of $Q_B$. Case (ii): $Q_CQ_BQ_A\xrightarrow{}111$, $XOR\xrightarrow{}0$. Similarly, $Q_A = 1$, due to the output of XOR. In the same way the previous states of $Q_B$ and $Q_A$ result in $Q_B = 1$ and $Q_C = 1$. Case (iii): $Q_CQ_BQ_A\xrightarrow{}110$, $XOR\xrightarrow{}1$. The previous output of XOR being $0$, drives the thru port of $M_1$ to $0$. The previous state of $Q_B$ and $Q_A$ results in $Q_B = 1$ and $Q_C = 1$. Case (iv): $Q_CQ_BQ_A\xrightarrow{}101$, $XOR\xrightarrow{}0$. The previous output of XOR being $1$, drives the thru port of $M_1$ to $1$. The previous state of $Q_B$ and $Q_A$ results in $Q_B = 0$ \& $Q_C = 1$. Case (v): $Q_CQ_BQ_A\xrightarrow{}010$, $XOR\xrightarrow{}0$. The previous output of XOR being 0, drives the thru port of $M_1$ to $0$. The previous state of $Q_B$ and $Q_A$ results in $Q_B = 1$ and $Q_C = 0$. Case (vi): $Q_CQ_BQ_A\xrightarrow{}100$, $XOR\xrightarrow{}1$. The previous output of XOR being 0, drives the thru port of $M_1$ to $0$. The previous state of $Q_B$ and $Q_A$ results in $Q_B = 0$ and $Q_C = 1$. Case (vii): $Q_CQ_BQ_A\xrightarrow{}001$, $XOR\xrightarrow{}1$. The previous output of XOR being 1, drives the thru port of $M_1$ to 1. The previous state of $Q_B$ and $Q_A$ results in $Q_B = 0$ and $Q_C = 0$. 

Similarly, considering the implementation of $4-$bit PRBS Generator, $M_1-M_4$ denote the $D$ flip-flops, whereas $M_5$ signifies the XOR gate. Here also, assuming the same initial bit pattern of $Q_DQ_CQ_BQ_A\xrightarrow{}0010$, the previous output of the XOR gate influences the output of $Q_A$ in the present clock cycle. Similarly, the previous output of $Q_D, Q_C$, and $Q_B$. Cases corresponding to each clock cycle are not discussed in detail. The main objective of depicting a 4-bit PRBS generator is due to its wide application in various processing architectures [39]. The architectural robustness of the sequence generator scheme allows reusing identical flip-flops according to the same operating principle, connected by some combinatorial network.

\begin{figure*}
    \centering
    \includegraphics[width = 18.1cm, height = 12.5cm]{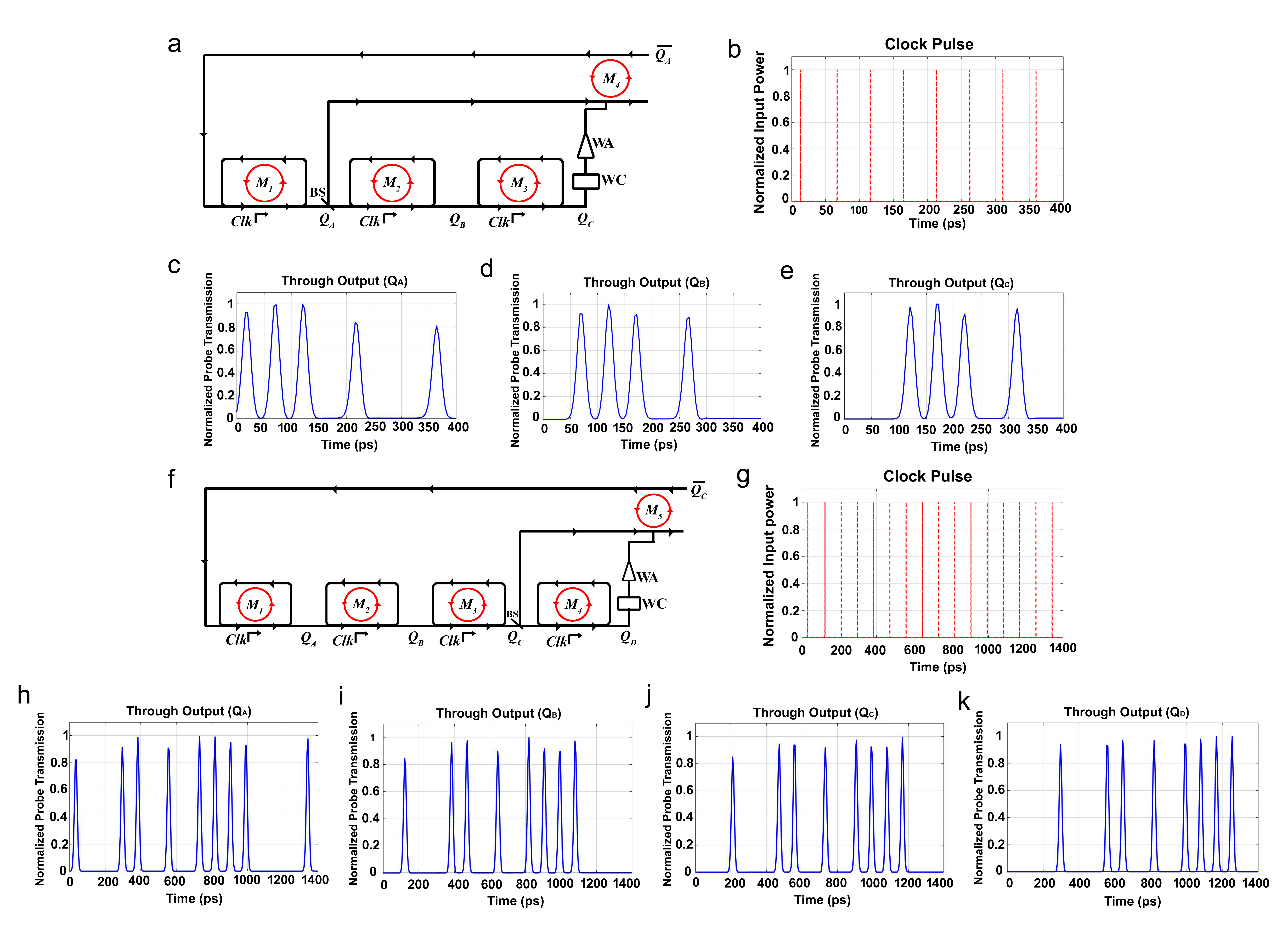}
    \caption{Design of all-optical Pseudo Random Binary Generator (PBRS), (a) Schematic of $3-$bit PBRS circuit, (b) high intensity pump signal, the clock pulses (red dashed line), (c-e) Simulated probe transmission with time (solid blue lines) of $3-$bit PRBS, corresponding to bits $Q_A$, $Q_B$, $Q_C$, (f) Schematic of $4-$bit PBRS circuit, (g) input pump signal clock pulses, (h-k) Simulated probe transmission of $4-$ bit PRBS circuit, corresponding to bits $Q_A$, $Q_B$, $Q_C$, $Q_D$. WC: Wavelength Converter, WA: Waveguide Amplifier.}
    \label{fig2}
\end{figure*}

\begin{table}[ht]
    \caption{TRUTH TABLE OF 4-BIT PRBS GENERATOR, WITH INITIAL SEQUENCE, $Q_{D}Q_{C}Q_{B}Q_{A}$ $\xrightarrow{}$ 0001}
    \centering
    \begin{tabular}{c|c|c|c|c|c}
    \hline\hline
    {Clock} & {$Q_D$} & {$Q_C$} & {$Q_B$} & {$Q_A$} & {$XOR$}\\
    {Cycle (Clk)} & (mW) & (mW) & (mW) & (mW) & Output (mW)\\
    \hline
    $0$ (initial) & 0 & 0 & 0 & 1(0.8) & 0\\
    $1^{st}$ & 0 & 0 & 1(0.9) & 0 & 0\\
    $2^{nd}$ & 0 & 1(0.9) & 0 & 0 & 1(1.3)\\
    $3^{rd}$ & 1(1.2) & 0 & 0 & 1(0.9) & 1(1.3)\\
    $4^{th}$ & 0 & 0 & 1(1.2) & 1(1.3) & 0\\
    $5^{th}$ & 0 & 1(1.1) & 1(1.3) & 0 & 1(1.2)\\
    $6^{th}$ & 1(1.2) & 1(1.1) & 0 & 1(1.0) & 0\\
    $7^{th}$ & 1(1.3) & 0 & 1(1.1) & 0 & 1(1.1)\\
    $8^{th}$ & 0 & 1(1.0) & 0 & 1(1.3) & 1(1.1)\\
    $9^{th}$ & 1(1.3) & 0 & 1(1.3) & 1(1.2) & 1(1.0)\\
    $10^{th}$ & 0 & 1(1.3) & 1(1.1) & 1(1.0) & 1(0.9)\\
    $11^{th}$ & 1(1.1) & 1(1.0) & 1(1.0) & 1(1.0) & 0\\
    $12^{th}$ & 1(1.2) & 1(1.0) & 1(1.2) & 0 & 0\\
    $13^{th}$ & 1(1.3) & 1(1.3) & 0 & 0 & 0\\
    $14^{th}$ & 1(1.3) & 0 & 0 & 0 & 1(0.9)\\
    $15^{th}$ (repeat) & 0 & 0 & 0 & 1(1.2) & 0\\
    \hline
    \hline
    \end{tabular}
    \label{table3}
\end{table}

\vspace{-1.5mm}
\subsection{4-bit Array Multiplier}
An ultrafast all-optical $4\times4$ array multiplier structure has been proposed in Fig. 3, in which Fig. 3(f)–(u) depicts its simulated response. The proposed design consists of $56$ MRR switches, where $M_i$ ($i$ = 1, 2, 3, 4, 5, 6, 7, 8, 15, 16, 17, 18, 31, 32, 33, 34) act as AND gates, $M_j$ ($j$ = 9, 10, 11, 12, 13, 14, 47, 48) acts as Half Adder (HA) switches, and $M_k$ ($k$ = 19, 20, 21, $\cdots$, 30, 35, 36, 37, $\cdots$, 46, 49, 50, 51 $\cdots$, 58) consists of Full Adder (FA) switches. $P_o$ and $P_7$ represent the least and the most significant bit of the multiplier output. $A_0A_1A_2A_3$ are considered as the probe input, shown in blue solid lines in Fig. 3(f)–(i). Input bits $B_0B_1B_2B_3$ are regarded as the pump inputs to the circuit, as shown in Fig. 3(j)–(m). The probe outputs obtained at the end of the AND gates $M_i(i=1, 2, 3, 4)$ are simultaneously given as a pump inputs to the subsequent half-adders (HAs) and full-adders (FAs). The output of the remaining AND gates are passed through a combination of WC and WA to increase intensity and change in wavelength ($\lambda_{probe} = 1568.75$ $nm\xrightarrow{}\lambda_{pump} = 1550.55$ $nm$) so that it functions as a pump for HAs and FAs. Since the probe, power is assumed to be $5\%$ of the pump power, the probe and pump can be efficiently interchanged by simply varying the intensities of the desired beams. Out of the 256 possible combinations resulting in the multiplication operation in Fig. 3(e), we report only the squared scenarios $(A_0A_1A_2A_3 = B_0B_1B_2B_3)$, corresponding to Table IV. 

Case (i): $A_0$ = $A_1$ = $A_2$ = $A_3$ = $B_0$ = $B_1$ = $B_2$ = $B_3$ = 0: no light is detected at any of the output ports. Case (ii): $A_0$ = $B_0$ = 1, $A_i$ = $B_i$ = $0$ $(i = 1,2,3)$: resulting in $P_o$ = 1 at 120 $ps$ [NPT $\sim95\%$ in Fig. 3(n)]. Case (iii): $A_1$ = $B_1$ = 1, $A_i$ = $B_i$ = $0$ $(i = 0,2,3)$: $A_1$ passes from $M_6$ to $M_{22}$ resulting in $P_2$ = 1. This corresponds to the simulated NPT $\sim95\%$ at 215 $ps$ as in Fig. 3(p). Case (iv): $A_o$ = $A_1$ = $B_o$ = $B_1$ = 1, $A_i$ = $B_i$ = $0$ $(i = 2,3)$: $A_o$ is switched by $M_1$ while $A_1$ passes from $M_2$ to $M_{19}$, finally to $M_{38}$ resulting in $P_o$ = $P_3$ = 1 at 310 $ps$. [NPT $\sim90\%$ as shown in Fig. 3(n), (q)]. 
\begin{table*}
\caption{GENERALIZED TRUTH TABLE FOR A $4\times4$ (4-bit) UNSIGNED ARRAY MULTIPLIER (A TOTAL OF 256 CASES ARE POSSIBLE)}
    \centering
    \setlength\tabcolsep{4pt}
    \begin{tabular}{l||c|c|c|c||c|c|c|c||c|c|c|c|c|c|c|c}
    \hline\hline
  Case & \multicolumn{4}{c}{Probe Input (mW)} & \multicolumn{4}{c}{Pump Input (mW)} & \multicolumn{8}{c}{Response from the Circuit (mW)}\\
    \cline{2-17}
    No. & $A_3$ & $A_2$ & $A_1$ & $A_0$ & $B_3$ & $B_2$ & $B_1$ & $B_0$ & $P_o$ & $P_1$ & $P_2$ & $P_3$ & $P_4$ & $P_5$ & $P_6$ & $P_7$ \\
     \hline
(i) & 0 & 0 & 0 & 0 & 0 & 0 & 0 & 0 & 0 & 0 & 0 & 0 & 0 & 0 & 0 & 0 \\
(ii) & 0 & 0 & 0 & 1(1.5) & 0 & 0 & 0 & 1(28) & 1(1.3) & 0 & 0 & 0 & 0 & 0 & 0 & 0 \\
(iii) & 0 & 0 & 1(1.5) & 0 & 0 & 0 & 1(28) & 0 & 0 & 0 & 1(1.3) & 0 & 0 & 0 & 0 & 0 \\
(iv) & 0 & 0 & 1(1.5) & 1(1.5) & 0 & 0 & 1(28) & 1(28) & 1(1.3) & 0 & 0 & 1(1.3) & 0 & 0 & 0 & 0 \\
(v) & 0 & 1(1.5) & 0 & 0 & 0 & 1(28) & 0 & 0 & 0 & 0 & 0 & 0 & 1(1.3) & 0 & 0 & 0 \\
(vi) & 0 & 1(1.5) & 0 & 1(1.5) & 0 & 1(28) & 0 & 1(28) & 1(1.2) & 0 & 0 & 1(1.2) & 1(1.3) & 0 & 0 & 0 \\
(vii) & 0 & 1(1.5) & 1(1.5) & 0 & 0 & 1(28) & 1(28) & 0 & 0 & 0 & 1(1.2) & 0 & 0 & 1(1.3) & 0 & 0 \\
(viii) & 0 & 1(1.5) & 1(1.5) & 1(1.5) & 0 & 1(28) & 1(28) & 1(28) & 1(1.2) & 0 & 0 & 0 & 1(1.3) & 1(1.3) & 0 & 0 \\
(ix) & 1(1.5) & 0 & 0 & 0 & 1(28) & 0 & 0 & 0 & 0 & 0 & 0 & 0 & 0 & 0 & 1(1.3) & 0 \\
(x) & 1(1.5) & 0 & 0 & 1(1.5) & 1(28) & 0 & 0 & 1(28) & 1(1.1) & 0 & 0 & 0 & 1(1.2) & 0 & 1(1.2) & 0 \\
(xi) & 1(1.5) & 0 & 1(1.5) & 0 & 1(28) & 0 & 1(28) & 0 & 0 & 0 & 1(1.2) & 0 & 0 & 1(1.2) & 1(1.2) & 0 \\
(xii) & 1(1.5) & 0 & 1(1.5) & 1(1.5) & 1(28) & 0 & 1(28) & 1(28) & 1(1.1) & 0 & 0 & 1(1.2) & 1(1.2) & 1(1.2) & 1(1.2) & 0 \\
(xiii) & 1(1.5) & 1(1.5) & 0 & 0 & 1(28) & 1(28) & 0 & 0 & 0 & 0 & 0 & 0 & 1(1.1) & 0 & 0 & 1(1.3) \\
(xiv) & 1(1.5) & 1(1.5) & 0 & 1(1.5) & 1(28) & 1(28) & 0 & 1(28) & 1(1.0) & 0 & 0 & 1(1.1) & 0 & 1(1.1) & 0 & 1(1.3) \\
(xv) & 1(1.5) & 1(1.5) & 1(1.5) & 0 & 1(28) & 1(28) & 1(28) & 0 & 0 & 0 & 1(1.1) & 0 & 0 & 0 & 1(1.2) & 1(1.2) \\
(xvi) & 1(1.5) & 1(1.5) & 1(1.5) & 1(1.5) & 1(28) & 1(28) & 1(28) & 1(28) & 1(0.9) & 0 & 0 & 0 & 0 & 1(1.0) & 1(1.1) & 1(1.1) \\
    \hline
    \hline
    \end{tabular}
    \label{table4}
\end{table*}

\begin{figure*}
    \centering
    \includegraphics[width = 18cm, height = 22.65cm]{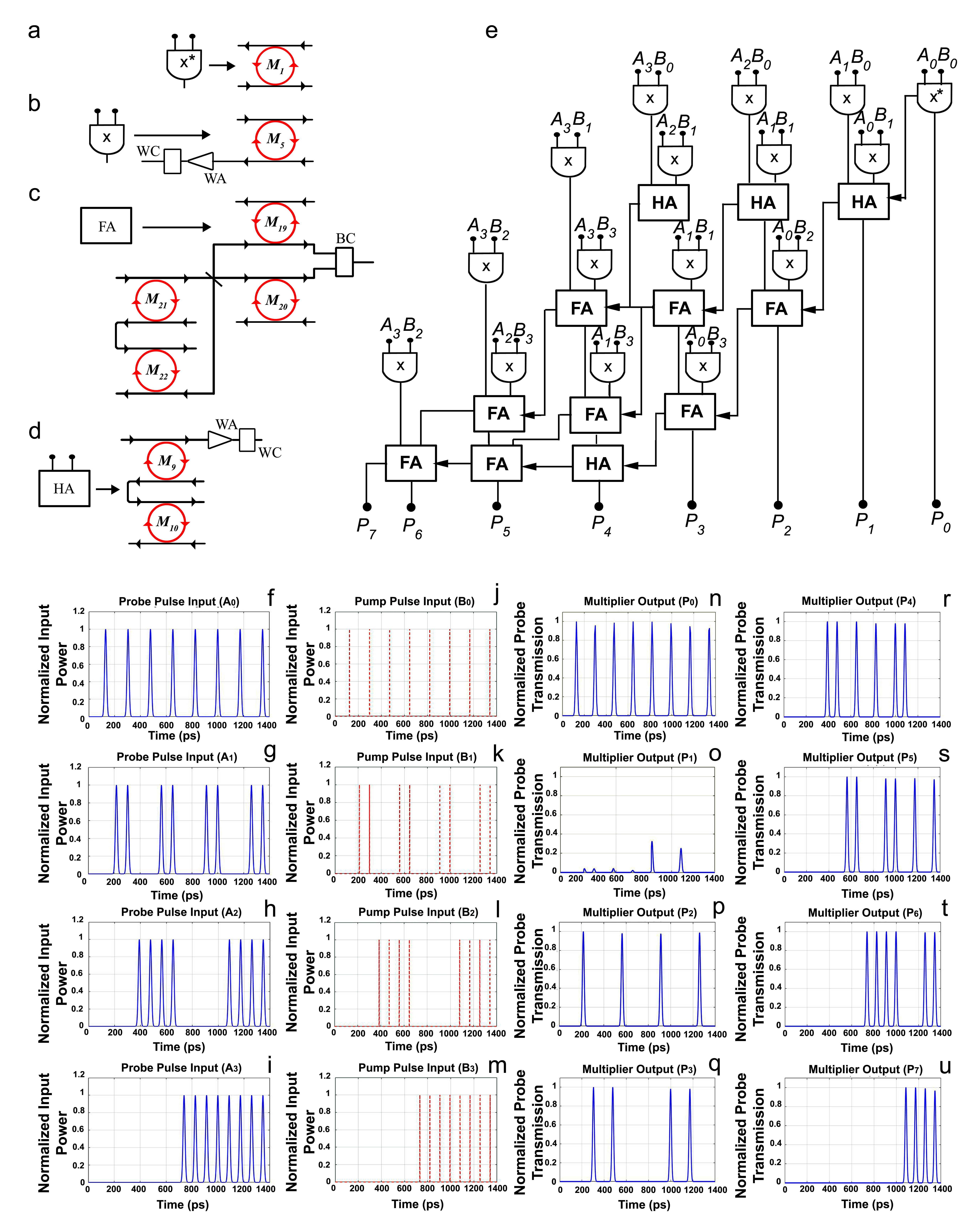}
    \caption{Design of all-optical $4\times4$-bit array Multiplier. Schematic of (a) and (b) AND Gates, (c) Full-Adder (FA), (d) Half-Adder (HA), (f), (g), (h), (i): low energy (weak) probe input bit $A_0$, $A_1$, $A_2$, $A_3$ to the circuit (solid blue lines). (j), (k), (l), (m): High intensity pump control input $B_0$, $B_1$, $B_2$, $B_3$ (dashed red line), (n-u): Simulated normalized probe transmission (NPT) with time of the proposed circuit, corresponding to Multiplier output $P_0$, $P_1$, $P_2$, $P_3$, $P_4$, $P_5$, $P_6$, $P_7$ (BC: Beam Combiner, WA: Waveguide Amplifier, WC: Wavelength Combiner)}
    \label{fig3}
\end{figure*}

Case (v): $A_2$ = $B_2$ = 1, $A_i$ = $B_i$ = $0$ $(i = 0,1,3)$: $A_2$ passes from $M_{17}$ to $M_{48}$ via $M_{30}$ and $M_{42}$ resulting in $P_4$ = 1 [NPT $\sim90\%$ at 405 $ps$ as shown in Fig. 4(n)]. Case (vi): $A_o$ = $B_o$ = $A_2$ = $B_2$ = 1, $A_i$ = $B_i$ = $0$ $(i = 1,3)$: $A_o$ is switched by $M_1$, and thus $P_o$ = 1 [NPT $\sim90\%$], while $A_2$ passes from $M_3$ to $M_{38}$, resulting in $P_3$ = 1 [NPT $\sim90\%$]. With the help of a BS, $A_o$ and $A_2$ passes from $M_{15}$ and $M_{17}$ to $M_{35}$ and $M_{48}$ ($A_o$ acts as a pump) resulting in $P_4$ = 1 [NPT $\sim90\%$] at 500 $ps$, as shown in Figs. 3(n),(q),(r). Case (vii): $A_1$ = $B_1$ = $A_2$ = $B_2$ = 1, $A_i$ = $B_i$ = $0$ $(i = 0,3)$: $A_1$ is passed from $M_6$ to $M_{12}$ resulting in $P_2$ = 1 [NPT $\sim85\%$] whereas $A_2$ is passed from $M_7$ and $M_{17}$ to $M_{52}$, resulting in $P_5$ = 1 [NPT $\sim95\%$] at 595 $ps$ as shown in Figs. 3(p),(s). Case (viii): $A_o$ = $B_o$ = $A_1$ = $B_1$ = $A_2$ = $B_2$ = 1, $A_3$ = $B_3$ = 0: $A_o$ is switched by $M_1$ resulting in $P_0$ = 1 [NPT $\sim 90\%$]. $A_1$ passes from $M_2$, $M_6$ and $M_{16}$ to $M_{48}$, resulting in $P_4$ = 1 [NPT $\sim90\%$]. $A_2$ passes from $M_3$, $M_7$ and $M_{17}$ to $M_{52}$, resulting in $P_5$ = 1 [NPT $\sim90\%$] at $690$ $ps$ as shown in Figs. 3(n),(r),(s). Case (ix): $A_3$ = $B_3$ = 1, $A_i$ = $B_i$ = $0$ $(i = 0,1,2)$: $A_3$ is passed from $M_{34}$ to $M_{55}$, giving $P_6$ = 1. This corresponds to NPT $\sim90\%$ at 785 $ps$, as shown in Fig. 4(p). Case (x): $A_o$ = $B_o$ = $A_3$ = $B_3$ = 1, $A_i$ = $B_i$ = $0$ $(i = 1,2)$: $A_o$ is switched by $M_1$ resulting in $P_o$ = 1 [NPT $\sim90\%$]. With the help of a BS, $A_o$ passes from $M_{31}$ to $M_{35}$, whereas $A_3$ passes from $M_4$ and $M_{34}$ to $M_{53}$, resulting in $P_4$ = $P_6$ = 1 [NPT $\sim90\%$ at 880 $ps$ as shown in Figs. 3(n),(r),(t)]. Case (xi): $A_1$ = $B_1$ = $A_3$ = $B_3$ = 1, $A_i$ = $B_i$ = $0$ $(i = 0,2)$: $A_1$ passes from $M_6$ to $M_{22}$, resulting in $P_2$ = 1 [NPT $\sim85\%$]. With the help of a BS, $A_3$ is switched from $M_4$ to $M_{52}$, resulting in $P_5$ = 1 [NPT $\sim 90\%$]. $A_1$ upon passing through a combination of WC and WA changes to a pump and gets switched from $M_{32}$ to $M_{52}$. Again, $A_{3}$ is switched by $M_{34}$ to act as a probe input to $M_{55}$, which results in $P_6$ = 1 [NPT $\sim 90\%$] at 975 $ps$, as shown in Figs. 4(p),(s),(t). Case (xii): $A_i$ = $B_i$ = 1 $(i = 0,1,3)$, $A_2$ = $B_2$ = 0: $A_o$ is switched by $M_1$ resulting in $P_o$ = 1 [NPT $\sim85\%$]. $A_1$ passes from $M_2$, $M_6$, and $M_{32}$ to $M_{38}$ and $M_{48}$, which results in $P_3$ = 1 [NPT $\sim85\%$] and $P_4$ = 1 [NPT $\sim85\%$]. $A_o$ can be modelled as a pump input to modulate input probe power. $A_3$ passes from $M_4$, $M_8$, $M_{34}$ to $M_{52}$ and $M_{53}$, which results in $P_5$ = 1 [NPT $\sim90\%$] and $P_6$ = 1 [NPT $\sim85\%$] at 1070 $ps$ as shown in  Figs. 3(n),(p),(q),(s),(t). Case (xiii): $A_2$ = $B_2$ = $A_3$ = $B_3$ = 1, $A_i$ = $B_i$ = $0$ $(i = 0,1)$: $A_2$ passes from $M_{17}$ and $M_{33}$ to $M_{48}$, and $M_{43}$. This results in $P_4$ = 1 [NPT $\sim 82\%$, Fig. 3(r)]. $A_3$ passes from $M_{18}$ to $M_{45}$. The output from $M_{34}$ and $M_{43}$ results in high output at the thru port of $M_{56}$ resulting in $P_7$ = 1 at 1165 $ps$ [NPT $\sim85\%$ as in Fig. 3(u)]. Case (xiv): $A_i$ = $B_i$ = 1 $(i = 0,2,3)$, $A_1$ = $B_1$ = 0: $A_o$ is switched by $M_1$ resulting in $P_o$ = 1 [NPT $\sim82\%$, Fig. 4(j)]. $A_3$ passes from $M_{4}$ to $M_{37}$ resulting in $P_3$ = 1 [NPT $\sim82\%$, Fig. 3(q)]. $A_2$ passes from $M_3$ to $M_{47}$ and with the help of a BS passes from $M_{17}$ to $M_{42}$ and acts as a pump pulse to $M_{51}$. $A_3$ passes from $M_{18}$ to $M_{51}$, resulting in $P_5$ = 1 [NPT $\sim85\%$, Fig. 3(s)]. $A_3$ also passes from $M_{34}$ to $M_{56}$, yielding $P_7$ = 1 at 1260 $ps$ [NPT $\sim85\%$, Fig. 3(u)]. Case (xv): $A_i$ = $B_i$ = 1 $(i = 1,2,3)$, $A_o$ = $B_o$ = 0: $A_1$ passes from $M_6$ to $M_{22}$ resulting in $P_2$ = 1 [NPT $\sim82\%$, Fig. 3(p)]. $A_2$ passes from $M_{7}$ to $M_{49}$ (where it acts as a pump). $A_2$ at input port of $M_{17}$ passes to $M_{54}$ and $M_{55}$ as a pump signal. $A_3$ passes from $M_{18}$ to $M_{53}$ and $M_{56}$ as a pump signal. Again, $A_3$ passes $M_{34}$ to reach the drop port ($P_d$) of $M_{55}$, resulting in $P_6$ = $P_7$ = 1 [NPT $\sim85\%$ at 1355 $ps$, as shown in Figs. 3(t),(u)]. Case (xvi): $A_i$ = $B_i$ = 1 $(i = 0,1,2,3)$: $A_o$ is switched by $M_1$ resulting in $P_o$ = 1 [NPT $\sim82\%$, Fig. 3(n)]. $A_1$ passes from $M_2$, $M_{6}$, $M_{16}$, and $M_{32}$ to $M_{19}$, $M_{23}$, and $M_{39}$, where they function as control pump signal. $A_2$ at the input port of $M_3$, $M_{7}$, $M_{17}$, and $M_{33}$ gets switched and amplified to act as a pump signal to $M_{23}$, $M_{27}$, and $M_{43}$. $A_3$ passes from $M_{18}$ to $M_{52}$, resulting in $P_5$ = 1 [NPT $\sim82\%$, Fig. 3(s)]. With the help of BS, $A_3$ at the input port of $M_{34}$ is switched to $M_{55}$, where $M_{53}$, $M_{54}$, and $M_{56}$ switches the probe input to port $P_6$ and $P_7$. This corresponds to $P_6$ = $P_7$ = 1, NPT $\sim85\%$ at 1450 $ps$, as shown in Figs. 3(t),(u). 

The 4-bit multiplier contains a number of rings. Design of such complex optical systems such requires a careful accounting of insertion losses, crosstalk, and other nonlinear effects [40-46]. The waveguide propagation losses can be considered $\sim2.7$ $dB/cm$ [43],while the waveguide crossing losses $\sim0.2$ $dB$. Due to the nature of the lithography process the roughness on the sidewall is unavoidable, this introduces a sidewall corrugation loss, which is considered the strongest effect for Silicon-On-Insulator (SOI) guides. Bends are effective parts of ring resonators, hence a bending loss of $0.01$ $dB/90^{o}$ can be considered for a $500$ $nm$ SOI waveguide [30]. This excess bend can be minimized by waveguide cross-section tuning or by engineering the bend shape [31-33, 40].

\subsection{4-bit Unsigned Divider}
The proposed architecture of the 4-bit divider is shown in Fig. 4(e), whereas Fig. 4(n)-(u) depicts the simulated response of the circuit. The design consists of $68$ MRR switches, where $M_i$ ($i$ = 9, 10, 11, 26, 27, 28, 43, 44, 45, 61, 62, 63, 64) acts as a 2-bit Multiplexer, $M_j$ ($j$ = 1, 2, 3, 4, 5, 6, 7, 8, 12, 13, 29, 30, 46, 47) acts as Half Substractor (HS) switches, $M_k$ ($k$ = 14, 15, 16, $\cdots$, 25, 31, 32, 33, $\cdots$, 42, 48, 49, 50, $\cdots$, 60) acts Full Substractor (FS) switches, and the remaining $M_{65}$, $M_{66}$, $M_{67}$, $M_{68}$ acts as NOT gates. $Q_0Q_1Q_2Q_3$ represents the Quotient bit, and $R_0R_1R_2R_3$ the remainder bit respectively. $A_0A_1A_2A_3$ is considered as the probe input, and $B_0B_1B_2B_3$ as the pump input, respectively. The probe output obtained at the end of the Multiplexers are passed onto the subsequent full substractor as a pump input. Output of the multiplexer is passed through a combination of WC and WA to increase its intensity and change its wavelength from probe to pump wavelength. 15 random trials have been considered to simulate the proposed circuit, which is shown in truth TABLE V. 

For Case (i): $A_0$ = $B_0$ = 1, $A_i$ = $B_i$ = 0 $(i = 1,2,3)$: $\Bar{A_3}$ passes from $M_{2}$ to $M_{65}$, resulting in $Q_0 = 1$ at 25 $ps$ [NPT $\sim95\%$ in Fig. 4(n)]. Case (ii): $A_0$ = $A_2$ = 1, $A_i$ = $0$ $(i = 1,3)$, $B_1$ = 1, $B_i$ = $0$ $(i = 0,2,3)$: $A_0$ is switched from $M_{48}$ to $M_{61}$, resulting in $R_0$ = 1 at 120 $ps$ [NPT $\sim95\%$, Fig. 4(r)]. $A_2$ passes from $M_{13}$ to $M_{42}$ and $M_{66}$ resulting in $Q_1$ = 1 at 120 $ps$ [NPT $\sim95\%$, Fig. 4(o)]. Case (iii): $A_0$ = $A_1$ = $A_3$ = $1$, $B_0$ = $B_1$ = $1$, $A_2$ = $B_i$ = 0, $(i = 2,3)$: $A_1$ passes from $M_{43}$ to $M_{62}$ resulting in $R_1$ = 1 [NPT $\sim95\%$ at 215 $ps$ as shown in Fig. 4(s)]. $A_3$ passes from $M_9$ to $M_{59}$ resulting in $Q_{0}$ = 1 [NPT $\sim95\%$ at 215 $ps$, Fig. 4(n)]. Whereas, $\Bar{A_{2}}$ passes from $M_{13}$ to $M_{66}$, resulting in $Q_1$ = 1 [NPT $\sim95\%$ at 215 $ps$, Fig. 4(o)]. Case (iv): $A_o$ = $A_3$ = 1, $B_1$ = 1, $A_i$ = $B_j$ = $0$ $(i = 1,2), (j = 0,2,3)$: $A_o$ is switched by $M_{61}$ while $A_1$ passes from $M_2$ to $M_{9}$, from which it is switched by $M_{67}$ resulting in $R_o$ = $Q_2$ = 1 at 310 $ps$. [NPT $\sim95\%$ as shown in Fig. 4(r),(p)]. Case (v): $A_0$ = $A_1$ = $A_2$ = $A_3$ = 1, $B_0$ = $B_1$ = $1$, $B_i$ = 0 $(i = 2,3)$: $A_3$ passes from $M_{9}$ to $M_{67}$, resulting in $Q_2$ = 1 [NPT $\sim90\%$ at 405 $ps$ as shown in Fig. 4(p)]. $A_2$ passes from $M_{26}$ to $M_{65}$, resulting in $Q_0$ = 1. This corresponds to a simulated NPT $\sim90\%$ at 405 $ps$ as shown in Fig. 4(n). Case (vi): $A_0$ = $A_2$ = $A_3$ = $1$, $B_1$ = $1$, $A_1$ = $B_i$ = 0, $(i = 0,2,3)$: $A_0$ is switched by $M_{61}$, and thus $R_o$ = 1 [NPT $\sim90\%$ at 500 $ps$, Fig. 4(r)], while $A_2$ passes from $M_{13}$ to $M_{26}$, further to $M_{37}$ and $M_{66}$, resulting in $Q_1$ = 1 [NPT $\sim90\%$ at 500 $ps$, Fig. 4(o)]. $A_3$ passes from $M_{2}$ to $M_{9}$ and further to $M_{25}$ ($A_2$ acts as a pump) resulting in $Q_2$ = 1 [NPT $\sim90\%$ at 500 $ps$, Fig. 4(p)]. Case (vii): $A_i = 1$, $(i = 0,1,2,3)$, $B_1 = 1$, $B_i = 0$, $(i = 0,1,2,3)$: $A_o$ is switched from $M_{48}$ to $M_{61}$ resulting in $R_o = 1$ [NPT $\sim85\%$ at 595 $ps$, Fig. 4(r)]. $A_3$ passes from $M_2$ to $M_{24}$ and $M_{25}$ further to $M_{67}$, resulting in $Q_2 = 1$ [NPT $\sim85\%$ at 595 $ps$, Fig. 4(p)]. $A_2$ passes from $M_{13}$ to $M_{26}$ and then to $M_{41}$ and $M_{42}$ resulting in $Q_1 = 1$ [NPT $\sim85\%$ at 595 $ps$, Fig. 4(o)]. $A_1$ passes from $M_{30}$ to $M_{43}$ and further to $M_{65}$, resulting in $Q_0 = 1$ [NPT $\sim85\%$ at 595 $ps$, Fig. 4(n)]. Case (viii): $A_3$ = $B_o$ = $B_1$ = 1, $A_i = 0$, $(i = 0,1,2)$, $B_i = 0$, $(i = 2,3)$: $\Bar{A_3}$ is switched from $M_2$ to $M_8$ and $M_7$ and further onto $M_{68}$ resulting in $Q_3 = 1$ [NPT $\sim95\%$ at 690 $ps$ as shown in Fig. 4(q)]. Case (ix): $A_o$ = $A_3$ = $B_o$ = 1, $A_i$ = $0$ $(i = 1,2)$, $B_i$ = 0 $(i = 1,2,3)$: $A_o$ passes from $M_{48}$ and $M_{47}$ to $M_{65}$, resulting in $Q_o$ = 1. [NPT $\sim82\%$ at 785 $ps$, as shown in Fig. 4(n)]. $A_3$ passes from $M_2$ to $M_7$ and $M_8$, and then to $M_{68}$, resulting in $Q_3 = 1$. [NPT $\sim85\%$ at 785 $ps$, as shown in Fig. 4(q)]. Case (x): $A_o$ = $A_1$ = $A_2$ = $B_2$ = 1, $A_3$ = $B_i$ = $0$ $(i = 0,1,3)$: $A_o$ is switched from $M_{48}$ to $M_{61}$, resulting in $R_o$ = 1 [NPT $\sim90\%$ at 880 $ps$, Fig. 4(r)]. $A_1$ is switched from $M_{30}$ to $M_{43}$, further down to $M_{62}$, resulting in $R_1 = 1$ [NPT $\sim85\%$ at 880 $ps$, Fig. 4(s)]. $A_2$ is switched from $M_{13}$ to $M_{26}$ and then to $M_{65}$, resulting in $Q_o = 1$ [NPT $\sim90\%$ at 880 $ps$, Fig. 4(n)]. Case (xi): $A_1$ = $A_3$ = $B_0$ = $B_2$ = 1, $A_i$ = $0$ $(i = 0,2)$, $B_i = 0$ $(i = 1,3)$: $A_1$ is passed from $M_{30}$ to $M_{42}$ and $M_{41}$ via $M_{34}$, resulting in $Q_1$ = 1 [NPT $\sim85\%$ at 975 $ps$, Fig. 4(o)] ($A_3$, $B_o$, $B_1$ works as a pump). Case (xii): $A_2$ = $A_3$ = $B_2$ = 1, $A_i$ = 0 $(i = 0,1)$, $B_i$ = 0 $(i = 0,1,3)$: $A_2$ passes from $M_2$ to $M_{66}$, resulting in $Q_o$ = 1 [NPT $\sim85\%$ at 1070 $ps$, Fig. 4(n)]. $A_3$ passes from $M_2$ to $M_{66}$, via $M_{9}$, $M_{27}$, and $M_{41}$ that results in $Q_1 = 1$ [NPT $\sim85\%$ at 1070 $ps$, Fig. 4(o)]. Case (xiii): $A_3$ = $B_1$ = 1, $A_i$ = $0$ $(i = 0,1,2)$, $B_i = 0$ $(i = 0,2,3)$: $A_3$ passes from $M_{2}$ to $M_{67}$ via $M_{9}$, $M_{17}$, $M_{24}$, and $M_{25}$. This results in $Q_2$ = 1 [NPT $\sim 82\%$ at 1165 $ps$, Fig. 4(p)]. Case (xiv): $A_o$ = $A_1$ = $B_1$ = 1, $A_i$ = 0 $(i = 2,3)$, $B_i$ = 0 $(i = 0,2,3)$: $A_o$ is switched from $M_{48}$ to $M_{61}$ resulting in $R_o$ = 1 [NPT $\sim80\%$ at 1260 $ps$, Fig. 4(j)]. $A_1$ passes from $M_{30}$ to $M_{65}$ via $M_{43}$, $M_{52}$, and $M_{59}$, resulting in $Q_o = 1$ [NPT $\sim80\%$ at 1260 $ps$, Fig. 4(n)]. Case (xv): $A_1$ = $A_2$ = $B_3$ = 1, $A_i$ = 0, $(i = 1,3)$, $B_i$ = 0 $(i = 0,1,2)$: $A_2$ is passed from $M_{13}$ to $M_{64}$ via $M_{26}$, $M_{36}$, $M_{45}$ and $M_{58}$, resulting in $R_3$ = 1 [NPT $\sim95\%$ at 1375 $ps$, Fig. 4(u)]. 

\begin{table*}
\caption{GENERALIZED TRUTH TABLE FOR A $4\times4$ (4-bit) UNSIGNED DIVIDER}
    \centering
    \setlength\tabcolsep{4pt}
    \begin{tabular}{l||c|c|c|c||c|c|c|c||c|c|c|c||c|c|c|c}
    \hline\hline
  Case & \multicolumn{4}{c}{Probe Input (mW)} & \multicolumn{4}{c}{Pump Input (mW)} & \multicolumn{8}{c}{Response from the Circuit (mW)}\\
    \cline{2-17}
    No. & $A_3$ & $A_2$ & $A_1$ & $A_0$ & $B_3$ & $B_2$ & $B_1$ & $B_0$ & $Q_3$ & $Q_2$ & $Q_1$ & $Q_0$ & $R_3$ & $R_2$ & $R_1$ & $R_0$ \\
     \hline
(i) & 0 & 0 & 0 & 1(1.5) & 0 & 0 & 0 & 1(28) & 0 & 0 & 0 & 1(1.3) & 0 & 0 & 0 & 0 \\
(ii) & 0 & 1(1.5) & 0 & 1(1.5) & 0 & 0 & 1(28) & 0 & 0 & 0 & 1(1.3) & 0 & 0 & 0 & 0 & 1(1.3) \\
(iii) & 1(1.5) & 0 & 1(1.5) & 1(1.5) & 0 & 0 & 1(28) & 1(28) & 0 & 0 & 1(1.3) & 1(1.3) & 0 & 0 & 1(1.3) & 0 \\
(iv) & 1(1.5) & 0 & 0 & 1(1.5) & 0 & 0 & 1(28) & 0 & 0 & 1(1.3) & 0 & 0 & 0 & 0 & 0 & 1(1.3) \\
(v) & 1(1.5) & 1(1.5) & 1(1.5) & 1(1.5) & 0 & 0 & 1(28) & 1(28) & 0 & 1(1.3) & 0 & 1(1.2) & 0 & 0 & 0 & 0 \\
(vi) & 1(1.5) & 1(1.5) & 0 & 1(1.5) & 0 & 0 & 1(28) & 0 & 0 & 1(1.2) & 1(1.2) & 0 & 0 & 0 & 0 & 1(1.3) \\
(vii) & 1(1.5) & 1(1.5) & 1(1.5) & 1(1.5) & 0 & 0 & 1(28) & 0 & 0 & 1(1.2) & 1(1.1) & 1(1.2) & 0 & 0 & 0 & 1(1.2) \\
(viii) & 1(1.5) & 0 & 0 & 0 & 0 & 0 & 0 & 1(28) & 1(1.3) & 0 & 0 & 0 & 0 & 0 & 0 & 0 \\
(ix) & 1(1.5) & 0 & 0 & 1(1.5) & 0 & 0 & 0 & 1(28) & 1(1.1) & 0 & 0 & 1(1.1) & 0 & 0 & 0 & 0 \\
(x) & 0 & 1(1.5) & 1(1.5) & 1(1.5) & 0 & 1(28) & 0 & 0 & 0 & 0 & 0 & 1(1.1) & 0 & 0 & 1(1.1) & 1(1.2) \\
(xi) & 1(1.5) & 0 & 1(1.5) & 0 & 0 & 1(28) & 0 & 1(28) & 0 & 0 & 1(1.0) & 0 & 0 & 0 & 0 & 0 \\
(xii) & 1(1.5) & 1(1.5) & 0 & 0 & 0 & 1(28) & 0 & 0 & 0 & 0 & 1(1.0) & 1(1.0) & 0 & 0 & 0 & 0 \\
(xiii) & 1(1.5) & 0 & 0 & 0 & 0 & 0 & 1(28) & 0 & 0 & 1(1.1) & 0 & 0 & 0 & 0 & 0 & 0 \\
(xiv) & 0 & 0 & 1(1.5) & 1(1.5) & 0 & 0 & 1(28) & 0 & 0 & 0 & 0 & 1(0.9) & 0 & 0 & 0 & 1(1.0) \\
(xv) & 0 & 1(1.5) & 1(1.5) & 0 & 1(28) & 0 & 0 & 0 & 0 & 0 & 0 & 0 & 1(1.3) & 0 & 0 & 0 \\
    \hline\hline
    \end{tabular}

    \label{table5}
\end{table*}

\begin{figure*}
    \centering
    \includegraphics[width = 18.1cm, height = 21.75cm]{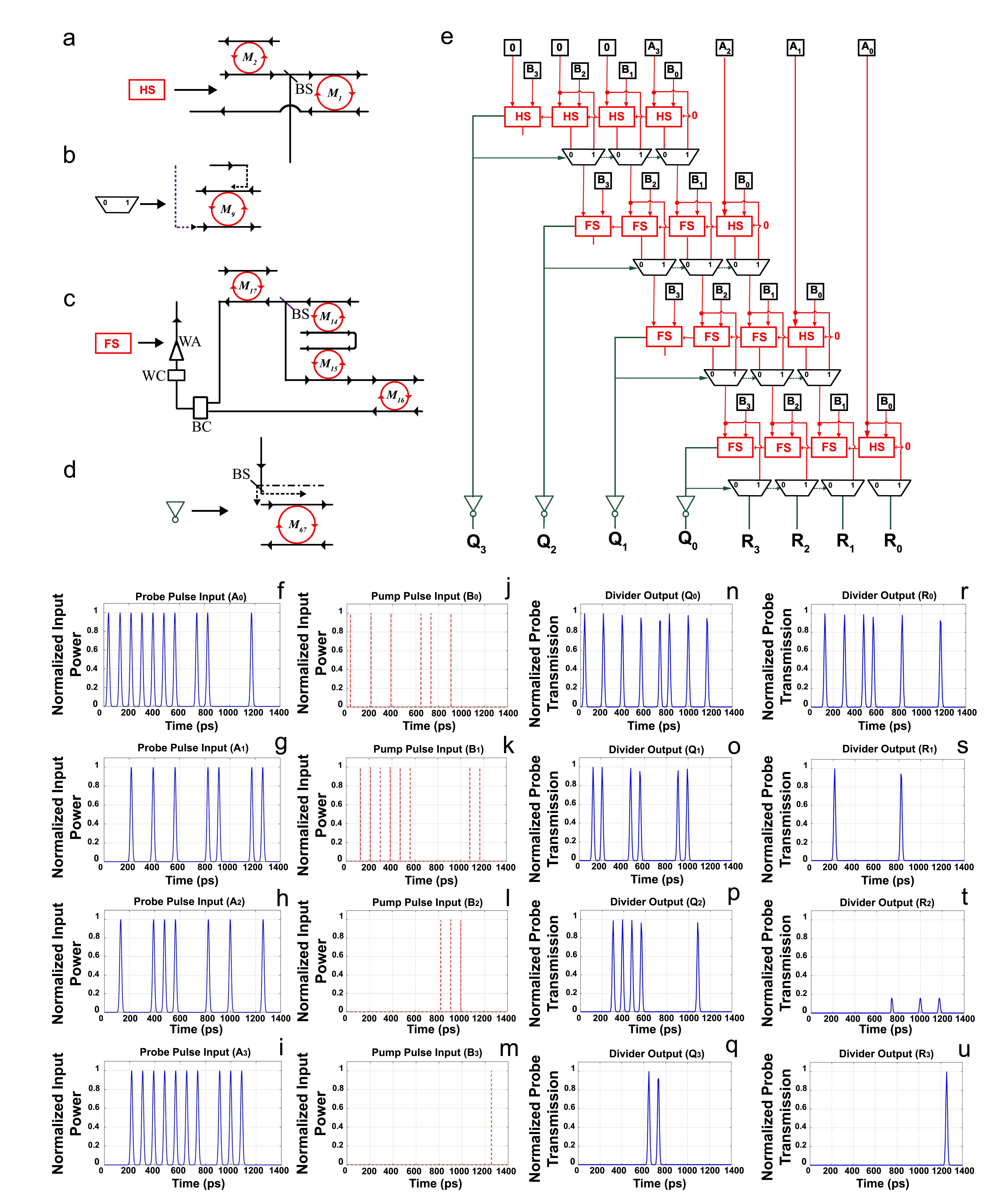}
    \caption{Schematic of the proposed all-optical $4\times4$-bit unsigned Divider, using $2\times2$ add-drop $Si$ MRRs. Realization of (a) Half-Substractor (HS) using optical switches, (b) $2\times1$ Multiplexer, (c) Full Substractor (FS), (d) NOT Gate. (f, g, h, i): low-energy (weak) probe input bit $A_0$, $A_1$, $A_2$, $A_3$ to the circuit (solid blue line). (j, k, l, m): High intensity pump control input $B_0$, $B_1$, $B_2$, $B_3$ (dashed red lines). Variation of normalized probe transmission (NPT) with time were studied for the cases corresponding to Truth Table VI. (n, o, p, q): Simulated NPT of the proposed circuit, corresponding to Quotient bits $Q_0$, $Q_1$, $Q_2$, $Q_3$. (r, s, t, u): Simulated NPT corresponding to remainder bits $R_0$, $R_1$, $R_2$, $R_3$. (BS: Beam Splitter, WA: Waveguide Amplifier, WC: Wavelength Combiner, BC: Beam Combiner)}
    \label{fig4}
\end{figure*}

MRRs are necessary due to small mode volume, high power density, and narrow spectral width. In ultrafast regimes, they can be considered as lumped oscillators that hold the best potential in terms of modulation energy per bit [47]. The proposed designs using SiMRRs are important for all-optical computing. Using multiple frequencies (control and probe) as logical inputs further reduces the number of switches with lower delays, enabling high bit-rate operation [48]. Since the input logic controls the state of each ring, all switches operate simultaneously, and the switching time does not accumulate. Hence, low latency and reconfigurability make the designs even more versatile.

The proposed designs of ultrafast all-optical PRBS generators are simpler and more compact using smaller number of SiMRRs with only one WA, besides also considering the thermo-optic effect, compared to the earlier reported designs [39, 65]. To the best of our knowledge, the proposed designs of ultrafast all-optical $4\times4$-bit multiplier and divider with SiMRRs are the first designs reported to date. Although the designs are simple, the proposed circuits involve WCs (wavelength converters), BSs (beam splitters), DCs (directional couplers) and WAs (waveguide amplifiers). WC has been demonstrated using four-wave mixing (FWM) [49]. SOAs are generally used for designing WCs due to their high conversion efficiency. Recently quantum-dot SOAs based on $InAs/InGaAs$ platform emerged as promising medium for efficient FWM-WC around telecom wavelengths [50]. To incorporate proposed designs to photonic integrated circuits (PICs), BSs with compact footprints are required [51–53]. The trade-off between footprint and insertion loss, makes it hard to design BSs for low-loss, compact logic operations. The similar analogy can be drawn for optical DCs. An optical DC based on traditional silicon wire waveguide has been demonstrated for optical interconnection in silicon chips [54]. The design shows a footprint of $40\times40$ $\mu m^2$ and an insertion loss of -$2.8$ dB. To realize designs in PIC, ultra-compact optical DCs are required [55]. In order to incorporate switches with optical interconnects fiber-to-fiber insertion and propagation losses have to be taken care, which also increase the price of PIC [56]. Achieving optical amplification using bulky EDFAs creates the same difficulty along with scalability and cascadability issues. In addition to SOAs, a wide range of on-chip ion-doped waveguide amplifiers have been reported that have advantages of gain, bandwidth, selectivity of host materials, low noise, insensitivity to polarization, good temperature stability, integrability and utility in multi-band optical communications [57]. Erbium-doped waveguide amplifier has recently been used to realize a lossless monolithically integrated re-configurable optical add-drop multiplexer [58].

The propagation of ultrashort optical pulses through nanowire waveguides induces dispersion, which is also detrimental for all-optical systems. Generally, group velocity dispersion (GVD) and third-order dispersion (TOD) result in distortion of ultrashort optical pulses, referred to as pump-pulse broadening effect [59]. Mathematically, the effects of fiber-optic dispersion in non-linear regime is tackled by expanding the mode-propagation constant $\beta$ and incorporating its effect on the frequency-dependent refractive index [59]. However, for the present simulations, this effect has not been considered due to the compact size of the circuits. The waveguide can also be engineered such that the effect of dispersion is mitigated. Recently many techniques that include, two-segment depletion mode modulation [60], dispersion tailoring [61], optimization of pump and probe wavelengths inducing non-reciprocity, i.e., having equal reciprocal group velocity [62], using dispersion compensators [63] or propagation under strong self-phase modulation have been reported [64].

\section{CONCLUSION}

All-optical switching in SiMRR was studied using time domain coupled-mode theory incorporating the effect of TPA-induced Free Carrier Dispersion (FCD), Free Carrier Absorption (FCA) and Thermo Optic Effect (TOE). A variable order Adams-Bashforth predictor corrector method was utilized to solve the coupled ODEs. All-optical implementation of complex logic operations, such as $3-$bit, and $4-$bit Pseudo Random Binary Sequence Generator, $4\times4$-bit all-optical multiplier and divider using $2\times2$ add-drop Silicon Microring resonators have been presented. The designs are general and can be implemented in both fiber-optic and integrated-optic formats. The designs were further optimized to realize ultrafast switching ($22$ $ps$), low-power operation ($28$ $mW$), and high modulation depth ($80\%$), operating at $45$ $Gb/s$. The proposed designs are important for realizing higher computing circuits due to CMOS–compatible computing, high Q-factor, tunability, compactness, cascadability and scalability. 
\end{document}